\documentclass[twocolumn,showpacs,amsmath,amssymb,pra,superscriptaddress]{revtex4-1}
\usepackage{amssymb}
\usepackage{amsmath}
\usepackage {longtable}
\usepackage{bm}

\usepackage{graphicx}
\newcommand{\ket}[1]{|#1\rangle}

\newcommand \be{\begin{equation}}
\newcommand \ee{\end{equation}}
\newcommand \bea{\begin{eqnarray}}
\newcommand \eea{\end{eqnarray}}
\newcommand \bse{\begin{subequations}}
\newcommand \ese{\end{subequations}}
\begin{document}
 \clubpenalty=10000 \widowpenalty=10000

\title{Adiabatic passage of radiofrequency-assisted F\"{o}rster resonances in Rydberg atoms for two-qubit gates and generation of Bell states}

\author{I.~I.~Beterov}
\email{beterov@isp.nsc.ru}
\affiliation {Rzhanov Institute of Semiconductor Physics SB RAS, 630090 Novosibirsk, Russia}
\affiliation {Novosibirsk State University, 630090 Novosibirsk, Russia}
\affiliation {Novosibirsk State Technical University, 630073 Novosibirsk, Russia}

\author{G.~N.~Hamzina}
\affiliation {Rzhanov Institute of Semiconductor Physics SB RAS, 630090 Novosibirsk, Russia}
\affiliation {Novosibirsk State Technical University, 630073 Novosibirsk, Russia}

\author{E.~A.~Yakshina}
\affiliation {Rzhanov Institute of Semiconductor Physics SB RAS, 630090 Novosibirsk, Russia}
\affiliation {Novosibirsk State University, 630090 Novosibirsk, Russia}

\author{D.~B.~Tretyakov}
\affiliation {Rzhanov Institute of Semiconductor Physics SB RAS, 630090 Novosibirsk, Russia}
\affiliation {Novosibirsk State University, 630090 Novosibirsk, Russia}
\author{V.~M.~Entin}
\affiliation {Rzhanov Institute of Semiconductor Physics SB RAS, 630090 Novosibirsk, Russia}
\affiliation {Novosibirsk State University, 630090 Novosibirsk, Russia}

\author{I.~I.~Ryabtsev}
\affiliation {Rzhanov Institute of Semiconductor Physics SB RAS, 630090 Novosibirsk, Russia}
\affiliation {Novosibirsk State University, 630090 Novosibirsk, Russia}

\begin{abstract}
High-fidelity entangled Bell states are of great interest in quantum physics. Entanglement of ultracold neutral atoms in two spatially separated optical dipole traps is promising for implementation of quantum computing and quantum simulation and for investigation of Bell states of material objects. We propose a new method to entangle two atoms via long-range Rydberg-Rydberg interaction. Alternatively to previous approaches, based on Rydberg blockade, we consider radiofrequency-assisted Stark-tuned F\"{o}rster resonances in Rb Rydberg atoms. To reduce the sensitivity of the fidelity of Bell states to the fluctuations of interatomic distance,  we propose to use the double adiabatic passage across the radiofrequency-assisted Stark-tuned F\"{o}rster resonances, which results in a deterministic phase shift of the two-atom state.

\end{abstract}
\maketitle

\section{Introduction}

\begin{center}
\begin{figure}[!t]
\center
\includegraphics[width=\columnwidth]{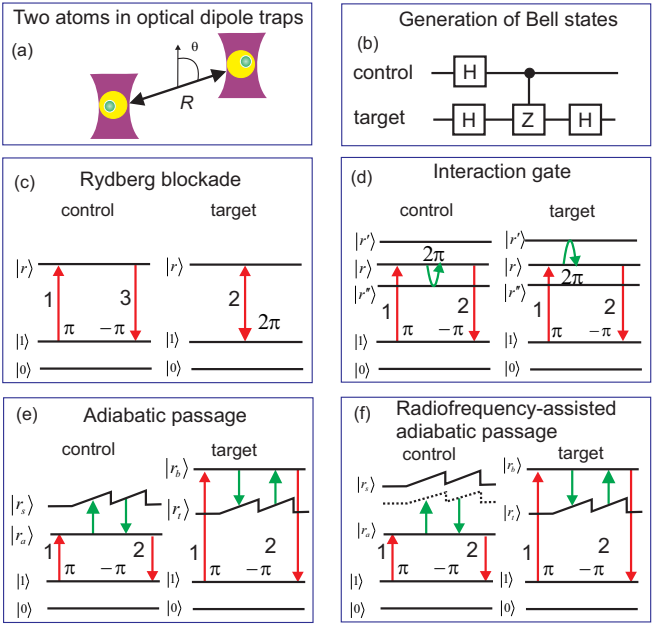}
\vspace{-.5cm}

\caption{
\label{Scheme}
(Color online) (a) Two Rydberg atoms in two optical dipole traps; (b) Scheme of generation of Bell states using Hadamard gates and CZ gate; (c) Scheme of CZ gate using Rydberg blockade; (d)  Scheme of  CZ gate using coherent dipole-dipole coupling at F\"{o}rster resonance; (e) Scheme of CZ gate using double adiabatic rapid passage across a Stark-tuned F\"{o}rster resonance. Two atoms (control and target) are excited to Rydberg states. An external time-dependent electric field shifts the energy levels of the Rydberg atoms so that the F\"{o}rster resonance is passed adiabatically two times. Then the atoms are de-excited to ground state. The phase shift is deterministically accumulated if both atoms are initially prepared in state $\ket{1}$; (f) Scheme of CZ gate using radiofrequency-assisted adiabatic rapid passage across a Stark-tuned F\"{o}rster resonance. The F\"{o}rster resonance is observed for the Floquet states, created by the radiofrequency electric field.
}

\end{figure}
\end{center}

High-fidelity entangled Bell states are of great interest in quantum physics~\cite{Bell1964}. Entanglement of ultracold neutral atoms in two spatially separated optical dipole traps~[see Fig.~\ref{Scheme}(a)]~\cite{Wilk2010, Isenhower2010} is promising for implementation of quantum computing and quantum simulation and for investigation of Bell states of material objects~\cite{Roos2004, Welte2017}. Arbitrary Bell states can be generated using a sequence of the Hadamard gates and a controlled-Z (CZ) gate applied to the control and target qubits as shown in Fig.~\ref{Scheme}(b)~\cite{Nielsen2011}.

The effect of Rydberg blockade~\cite{Lukin2001} has been successfully used for entanglement of two atoms~\cite{Wilk2010} and implementation of two-qubit gates in a scalable quantum register~\cite{Isenhower2010, Xia2015,Saffman2010}. Scheme of CZ gate based on Rydberg blockade is shown in Fig.~\ref{Scheme}(c). If the control qubit is excited to the Rydberg state by pulse~1, the excitation of the target qubit by pulse~2 will be blocked~\cite{Isenhower2010}. Otherwise, the $2\pi$ pulse~2, which acts on the target qubit, will shift the phase of the collective two-atom state by $\pi$.  Despite the simplicity of this scheme, high-fidelity two-qubit gates have not been implemented yet~\cite{Xia2015}. One of the possible obstacles is a complex structure of collective Rydberg states of two multi-level atoms which may result in blockade breakdown~\cite{Derevianko2015}.

Alternatively, the interaction of two temporarily excited Rydberg atoms can be used for implementation of controlled-phase gates via phase shifts, which are induced by the weak Rydberg-Rydberg interactions~\cite{Jaksch2000}, or due to their coherent coupling at a F\"{o}rster resonance~\cite{Ravets2014}. The scheme of CZ gate based on a Stark-tuned F\"{o}rster resonance is shown in Fig.~\ref{Scheme}(d)~\cite{Ravets2014, Ryabtsev2005}. Two trapped atoms are excited into Rydberg states $\ket{r}$ which are then tuned by the external electric field midway between two other Rydberg states of the opposite parity $\ket{r'}$ and $\ket{r''}$ (F\"{o}rster resonance)~\cite{Safinya1981, Anderson1998, Mourachko1998, Westermann2006, Nipper2012, Richards2016, Kondo2016, Pelle2016, Mandoki2016, Browaeys2016}.  The interaction between Rydberg atoms  should be sufficiently small to avoid Rydberg blockade during laser excitation. At F\"{o}rster resonance, Rydberg interaction coherently drives the transition $\ket{rr} \to \ket{r'r''}$, leading to the $\pi$ phase shift of the collective state $\ket{rr}$ after the end of single Rabi-like population oscillation.  In contrast to Rydberg blockade, this approach does not require strong Rydberg-Rydberg interaction, but it suffers from the sensitivity to the fluctuations of the interaction energy due to fluctuations of the interatomic distance.

Recently we proposed the schemes of CZ and CNOT gates based on the double adiabatic passage across the Stark-tuned F\"{o}rster resonance leading to a deterministic phase shift of the two-atom state~\cite{Beterov2016a}. Scheme of two-qubit CZ gate is shown in Fig.~\ref{Scheme}(e). Two atoms are excited into Rydberg states $\ket{r_a}$ and $\ket{r_b}$ which can be either different or identical. The time-dependent electric field adjusts the Rydberg energy levels in such a way that the energy difference between Rydberg state $\ket{r_a}$ and $\ket{r_s}$ becomes equal to the energy difference between Rydberg states $\ket{r'}$ and $\ket{r'''}$  twice in time. The states $\ket{r}$ and $\ket{r''}$ must be of the opposite parity, as well as the states $\ket{r_b}$ and $\ket{r_t}$.  Rydberg-Rydberg interaction leads to the adiabatic transfer of the population of the collective two-atom states $\ket{r_a r_b}\to \ket{r_s r_t}\to\ket{r_a r_b} $. After the end of the double adiabatic passage the system returns to the initial two-atom state $\ket{r_a r_b}$ with a deterministic $\pi$ phase shift added.

Implementation of this scheme requires isolated F\"{o}rster resonances for high-lying Rydberg states, which should have substantially long lifetimes in order to reduce decoherence during adiabatic passage~\cite{Beterov2009}. Therefore in our previous work~\cite{Beterov2016a} we considered only Cs atoms, where we have found a suitable resonance.

Radiofrequency-assisted or microwave-assisted F\"{o}rster resonances can be used to induce the "inaccessible" resonances which cannot be tuned by the dc electric field alone~\cite{Tauschinsky2008, Tretyakov2014, Yakshina2016}.  Additional isolated Stark-tuned resonances can thus be found. In this work we consider generation of Bell states using CZ gate, based on the double adiabatic passage of radiofrequency-assisted F\"{o}rster resonances, as shown in Fig.~\ref{Scheme}(f). This scheme is equivalent to Fig.~\ref{Scheme}(e), but the "inaccessible" F\"{o}rster resonance is observed for the additional Floquet states, which are induced by an external radiofrequency (rf) electric field~\cite{Tauschinsky2008, Tretyakov2014, Yakshina2016}.

This paper is organized as follows. In Sec.~II we present a theory of rf-assisted adiabatic passage in a two-level quantum system.  Sec.~III is devoted to Stark-tuned and rf-assisted F\"{o}rster resonances for Rb Rydberg atoms taking into account the complex structure of Rydberg energy levels and their behavior in the external electric field. In Sec.~IV we discuss the results of calculation of infidelities of Bell states taking into account finite lifetimes of Rydberg states and  the fluctuations of the interatomic distance.

\section{Theory of rf-assisted adiabatic passage}

The Hamiltonian for a two-level system with states $\ket{1}$ and $\ket{2}$ is written as 
\be
\label{eq1}
\hat{H}=\hbar \left(\begin{array}{cc} {0} & {V} \\ {V} & {\delta \left(t\right)} \end{array}\right).
\ee  
\noindent Here $\hbar V$ is a coupling energy and $\hbar \delta \left(t\right)$ is a time-dependent energy defect. Using the Schr\"{o}dinger equation, we write the equations for the probability amplitudes: 

\be
\label{eq2}
\begin{array}{l} {i\dot{c}_{1} =Vc_{2}}, \\ {i\dot{c}_{2} =\delta \left(t\right)c_{2} +Vc_{1}.} \end{array}
\ee

To study rf-assisted adiadatic passage, we write the time-dependent detuning as following:

\be
\label{eq3}
\delta \left(t\right)=\delta '\left(t\right)+\delta _{0} +A\cos \left(\omega _{\mathrm{rf}} t\right).  
\ee

\noindent Here $\delta '\left(t\right)$ is a time-dependent function, which is slowly varied around zero, $\delta _{0} =\mathrm{const}$ is the detuning for rf-assisted resonance, $\omega _{\mathrm{rf}}$ is the frequency of the  rf field and \textit{A} is the amplitude of modulation. After replacement

\be
\label{eq4}
c_{2} =\tilde{c}_{2} \exp \left[-i\frac{A}{\omega _{\mathrm{rf}} } \sin \left(\omega_{\mathrm{rf}} t\right)\right]
\ee
\noindent we obtain from Eq.~(\ref{eq2}):
\be
\label{eq5}
\begin{array}{l} 
{i\dot{c}_{1} =V\exp \left[-i\frac{A}{\omega _{\mathrm{rf}} } \sin \left(\omega _{\mathrm{rf}} t\right)\right]\tilde{c}_{2} }, \\ 
{i\dot{\tilde{c}}_{2} =\left[\delta '\left(t\right)+\delta _{0} \right]\tilde{c}_{2} +V\exp \left[i\frac{A}{\omega _{rf} } \sin \left(\omega _{\mathrm{rf}} t\right)\right]c_{1}. } \end{array}. 
\ee

Now we use the expansion~\cite{Abramowitz1972}

\be
\label{eq6}
\exp \left[i\frac{A}{\omega _{\mathrm{rf}} } \sin \left(\omega _{\mathrm{rf}} t\right)\right]=\sum _{m=-\infty }^{\infty }J_{m} \left(\frac{A}{\omega _{\mathrm{rf}} } \right) e^{ i m\omega _{\mathrm{rf}} t}.
\ee

\noindent and replacement $\tilde{c}_{2} =c'_{2} \exp \left(-i\omega _{\mathrm{rf}} t\right)$. Here $J_{m} \left(z\right)$ is a Bessel function of the first kind~\cite{Abramowitz1972}. By neglecting rapidly oscillating terms $\sim \exp \left(-im\omega _{\mathrm{rf}} t\right)$,  $m\ne 0$, we rewrite Eq.~(\ref{eq5}) as following:

\be
\label{eq7}
\begin{array}{l} 
{i\dot{c}_{1} =VJ_{-1} \left(\frac{A}{\omega_{\mathrm{rf}}}\right) c'_2}, \\ 
{i\dot{c'_{2} =\left[\delta '\left(t\right)+\delta _{0} -\omega _{\mathrm{rf}} \right]c'_{2} +VJ_{-1} \left(\frac{A}{\omega _{\mathrm{rf}} } \right)c_{1} }}. \end{array} 
\ee

If $\delta _{0} =\omega_{\mathrm{rf}}$, we can write the equivalent Hamiltonian as 

\be
\label{eq8}
\hat{H}'=\hbar \left(\begin{array}{cc} {0} & {VJ_{-1} \left({A \mathord{\left/{\vphantom{A \omega_{\mathrm{rf}} }}\right.\kern-\nulldelimiterspace} \omega _{rf} } \right)} \\ {VJ_{-1} \left({A \mathord{\left/{\vphantom{A \omega _{rf} }}\right.\kern-\nulldelimiterspace} \omega _{\mathrm{rf}} } \right)} & {\delta '\left(t\right)} \end{array}\right).
\ee

\begin{center}
\begin{figure}[!t]
\center
\includegraphics[width=\columnwidth]{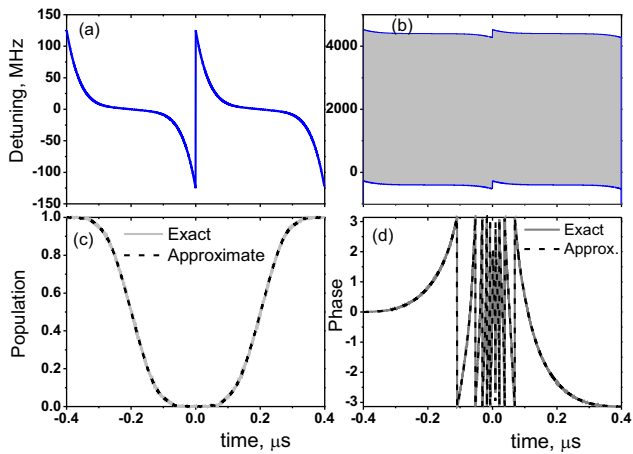}
\vspace{-.5cm}

\caption{
\label{Theory}
(Color online)(a) Nonlinear time-dependent detuning $\delta '\left(t\right)$ for adiabatic passage of Stark-tuned F\"{o}rster resonances; (b) Time-dependent detuning $\delta \left(t\right)$ for rf-assisted adiabatic rapid passage (rapid oscillations in the shaded region are not resolved); (c) Time dependence of the population of the initial state for rf-assisted adiabatic rapid passage calculated with the Hamiltonian from Eq.~(\ref{eq1}) (exact) and the Hamiltonian from Eq.~(\ref{eq8}) (approximate); (d) Time dependence of the phase of the initial state for rf-assisted adiabatic rapid passage calculated with the Hamiltonian from Eq.~(\ref{eq1}) (exact) and the Hamiltonian from Eq.~(\ref{eq8}) (approximate).
}

\end{figure}
\end{center}

Similarly, if $\delta _{0} =-\omega_{\mathrm{rf}} $, by replacement $\tilde{c}_{2} =c''_{2} \exp \left(i\omega_{\mathrm{rf}} t\right)$ in Eq.~((\ref{eq5}) we obtain 
\be
\label{eq9}
\hat{H}''=\hbar \left(\begin{array}{cc} {0} & {VJ_{1} \left({A \mathord{\left/{\vphantom{A \omega_{\mathrm{rf}} }}\right.\kern-\nulldelimiterspace} \omega _{\mathrm{rf}} } \right)} \\ {VJ_{1} \left({A \mathord{\left/{\vphantom{A \omega _{\mathrm{rf}} }}\right.\kern-\nulldelimiterspace} \omega_{\mathrm{rf}} } \right)} & {\delta '\left(t\right)} \end{array}\right).
\ee

In our previous work~\cite{Beterov2016} we considered adiabatic passage of the F\"{o}rster resonances with constant interaction energy. To achieve high fidelity of population transfer and deterministic phase accumulation, we used a double adiabatic sequence with nonlinear time-dependent detuning: 

\be
\label{eq10}
\delta'_{j} \left(t\right)=s_{1} \left(t-t_{m} \right)+s_{2} \left(t-t_{m} \right)^{5},  
\ee

\noindent as shown in Fig.~\ref{Theory}(a). Here the exact resonance occurs at times $t_{1} =-0.2\; \mu s$ and $t_{2} =0.2\; \mu s$, the parameters being ${s_{1}  \mathord{\left/{\vphantom{s_{1}  2\pi }}\right.\kern-\nulldelimiterspace} 2\pi } =-51\;{{\rm MHz} \mathord{\left/{\vphantom{{\rm MHz} \mu s}}\right.\kern-\nulldelimiterspace} \mu s} $ and ${s_{2}  \mathord{\left/{\vphantom{s_{2}  2\pi }}\right.\kern-\nulldelimiterspace} 2\pi } =-360\; {{\rm GHz} \mathord{\left/{\vphantom{{\rm GHz} \mu s}}\right.\kern-\nulldelimiterspace} \mu s} ^{5} $. 

Now we consider rf-assisted adiabatic passage with the time-dependent detuning from Eq.~(\ref{eq3}) with ${\delta_0/(2 \pi)=\omega _{\mathrm{rf}}  \mathord{\left/{\vphantom{\omega _{\mathrm{rf}}  2\pi }}\right.\kern-\nulldelimiterspace} 2\pi }=2\; {\rm GHz}$ and ${A \mathord{\left/{\vphantom{A 2\pi }}\right.\kern-\nulldelimiterspace} 2\pi } =2.4\; {\rm GHz}$, as shown in Fig.~\ref{Theory}(b). The comparison of the calculated time dependence of the population and phase of the initial state $\ket{1}$ of the system with the Hamiltonians from Eq.~(\ref{eq1})(exact) and Eq.~(\ref{eq9})(approximate) is shown in Figs.~\ref{Theory}(c) and \ref{Theory}(d), respectively. The jumps in Fig.~\ref{Theory}(d) and subsequent figures appear because the phase is defined in the interval $(-\pi,\pi)$. The interaction energy is ${V \mathord{\left/{\vphantom{V 2\pi }}\right.\kern-\nulldelimiterspace} 2\pi } =4.5$~MHz. Good agreement between the exact and approximate calculations is observed. This shows that the technique of deterministic phase accumulation via double adiabatic passage with nonlinear time-dependent detuning from the resonance, developed in our previous work, can also be applied to the rf-assisted F\"{o}rster resonances.

\section{Stark-tuned and rf-assisted F\"{o}rster resonances for Rb Rydberg atoms}

\begin{center}
\begin{figure}[!t]
\center
\includegraphics[width=\columnwidth]{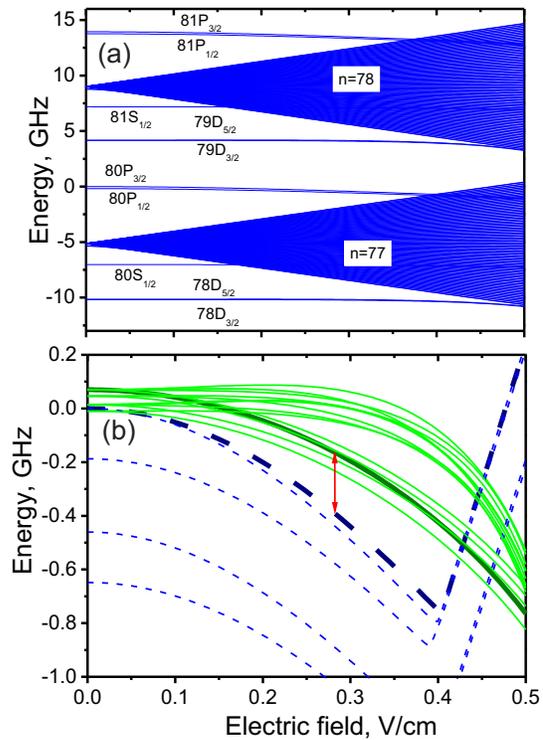}
\vspace{-.5cm}
\caption{
\label{Stark}
(Color online) (a) Stark diagram for Rb Rydberg state with $\left|m_j \right|=1/2$. The $80P_{3/2}$ state is selected as zero energy level. (b) The dependence of the energies of the collective states for $\ket{60P, \,80P}\to \ket{59D, \,78D}$ F\"{o}rster resonance on the electric field. The position of the selected rf-assisted F\"{o}rster resonance is indicated by the arrow.}
\end{figure}
\end{center}

Stark-tuned F\"{o}rster resonance required for the implementation of the proposed schemes must meet the following criteria~\cite{Beterov2016a}: (i) the lifetimes of Rydberg states must be sufficiently long to avoid the decay of coherence during the gate operation due to spontaneous and blackbody radiation (BBR) induced transitions; (ii) initial F\"{o}rster energy defect must be sufficiently large to allow for rapid turning off the interaction between atoms at the beginning and the end of the adiabatic passage; (iii) selected interaction channel must be well isolated from the other channels to avoid break-up or dephasing of the adiabatic population transfer. As a rule, such resonances are difficult to find.

The numerically calculated Stark diagram for Rb Rydberg states with $\left|m_j \right|=1/2$ is shown in Fig.~\ref{Stark}(a). The dc electric field is aligned along the \textit{z} axis. The 80$P_{3/2}$ state is selected as zero energy level. We have calculated the radial matrix elements using the quasiclassical approximation~\cite{Kaulakys1995} and the method of quantum defects~\cite{Cano2012}. The $\ket{S_{1/2};S_{1/2}}\to \ket{P_{1/2};P_{1/2}}$ interaction channel, used in our previous work~\cite{Beterov2016a}, has no splitting in the dc electric field. However, the anticrossing with the hydrogen-like manifold for 80\textit{S} state occurs in the electric field $E\sim 0.17$~V/cm which is too small for efficient implementation of the rf-assited adiabatic passage. The \textit{P} and \textit{D} states are sufficiently far from the hydrogenic manifold, but the F\"{o}rster resonances involving these states are split in the electric field, which makes it difficult to find a suitable isolated resonance. Therefore we exploit the angular dependence of the dipole-dipole interaction of Rydberg atoms at F\"{o}rster resonances which has been studied in Ref.~\cite{Ravets2015}.

The electric dipole-dipole interaction between two atoms A and B is described by the operator 

\be 
\label{eq11} 
\begin{array}{l} {\hat{V}_{dd} =\dfrac{1}{4\pi \varepsilon _{0} R^{3} } \times } \\ {\left[S_{1} \left(\theta \right)\left(\hat{d}_{A+} \hat{d}_{B-} +\hat{d}_{A-} \hat{d}_{B+} +2\hat{d}_{Az} \hat{d}_{Bz} \right)+\right. } \\ {+S_{2} \left(\theta \right)\left(\hat{d}_{A+} \hat{d}_{Bz} -\hat{d}_{A-} \hat{d}_{Bz} +\hat{d}_{Az} \hat{d}_{B+} -\hat{d}_{Az} \hat{d}_{B-} \right)-} \\ {\left. -S_{3} \left(\theta \right)\left(\hat{d}_{A+} \hat{d}_{B+} +\hat{d}_{A-} \hat{d}_{B-} \right)\right].} \end{array} 
\ee

\noindent Here $\hat{d}_{k,\pm } ={\mp \left(\hat{d}_{k,x} \pm i\hat{d}_{k,y} \right) \mathord{\left/{\vphantom{\mp \left(\hat{d}_{k,x} \pm i\hat{d}_{k,y} \right) \sqrt{2} }}\right.\kern-\nulldelimiterspace} \sqrt{2} }$ are the components of the dipole operator in the spherical basis and the angular prefactors are 

\be
\label{eq12} 
\begin{array}{l} 
{S_{1} \left(\theta \right)=\dfrac{1-3\cos^{2} \left(\theta \right)}{2}}, \\ 
{S_{2} \left(\theta \right)=\dfrac{3\sin \left(\theta \right)\cos \left(\theta \right)}{\sqrt{2} } }, \\ 
{S_{3} \left(\theta \right)=\dfrac{3\sin ^{2} \left(\theta \right)}{2} }. \end{array}
\ee

The operator $\hat{V}_{dd}$ couples two-atom states where the total magnetic quantum number $M=m_{1} +m_{2}$ changes by $\Delta M=0$, $\pm 1$ and $\pm 2$. For $\theta=\pi/2$ only the states with $\Delta M=0$ and $\Delta M=2$ are coupled. 

The matrix element of the $\hat{V}_{dd} $ operator for transition between the collective two-atom states $\ket{ r_a r_b} \to \ket{ r_s r_t} $,  where for each atomic state $\ket{r} =\ket{ nljm_{j}}$ \textit{n} is a principal quantum number, \textit{l} is the angular moment,  \textit{j} is the total moment and $m_j$ is the projection of the total moment, is expressed as~\cite{Varshalovich1998,Kamenski2017}

\be
\label{eq13} 
\begin{array}{l} {{\left\langle n_{s} m_{s} l_{s} j_{s} ;n_{t} m_{t} l_{t} j_{t}  \right|} \hat{V}_{dd} {\left| n_{a} m_{a} l_{a} j_{a} ;n_{b} m_{b} l_{b} j_{b}  \right\rangle} =} \\ 
{=\dfrac{e^{2} }{4\pi \varepsilon _{0} R^{3} } \left\{A_{1} \left(\theta \right)\left[2C_{j_{a} m_{a} 10}^{j_{s} m_{s} } C_{j_{b} m_{b} 10}^{j_{t} m_{t} } \right. +\right. } \\
 {\left. C_{j_{a} m_{a} 11}^{j_{s} m_{s} } C_{j_{b} m_{b} 1-1}^{j_{t} m_{t} } +C_{j_{a} m_{a} 1-1}^{j_{s} m_{s} } C_{j_{b} m_{b} 11}^{j_{t} m_{t} } \right]+} \\ 
{A_{2} \left(\theta \right)\left[\left(C_{j_{a} m_{a} 11}^{j_{s} m_{s} } -C_{j_{a} m_{a} 1-1}^{j_{s} m_{s} } \right)C_{j_{b} m_{b} 10}^{j_{t} m_{t} } +\right. } \\ 
{\left. C_{j_{a} m_{a} 10}^{j_{s} m_{s} } \left(C_{j_{b} m_{b} 11}^{j_{t} m_{t} } -C_{j_{b} m_{b} 1-1}^{j_{t} m_{t} } \right)\right]-} \\ 
{\left.A_{3} \left(\theta \right)\left[C_{j_{a} m_{a} 11}^{j_{s} m_{s} } C_{j_{b} m_{b} 11}^{j_{t} m_{t} } +C_{j_{a} m_{a} 1-1}^{j_{s} m_{s} } C_{j_{b} m_{b} 1-1}^{j_{t} m_{t} } \right]\right\}\times } \\ 
{\sqrt{\max \left(l_{a} ,l_{s} \right)} \sqrt{\max \left(l_{b} ,l_{t} \right)} \sqrt{\left(2j_{a} +1\right)\left(2j_{b} +1\right)} \times } \\ {\times \left\{\begin{array}{ccc} {l_{a} } & {1/2} & {j_{a} } \\ 
 \hline {j_{s} } & {1} & {l_{s} } \end{array}\right\}\left\{\begin{array}{ccc} {l_{b} } & {1/2} & {j_{b} } \\  
\hline {j_{t} } & {1} & {l_{b} } \end{array}\right\} \left(-1\right)^{l_{s} +\frac{l_{a} +l_{s} +1}{2} }\times } 
\\ 
{\times \left(-1\right)^{l_{t} +\frac{l_{b} +l_{t} +1}{2} } \left(-1\right)^{j_{a} +j_{b} } R_{n_{a} l_{a} }^{n_{s} l_{s} } R_{n_{b} l_{b} }^{n_{t}l_{t} }}. \end{array} 
\ee 

\noindent Here \textit{R} is the interatomic distance and $R_{n_{a} l_{a} }^{n_{s} l_{s} }$ and $R_{n_{b} l_{b} }^{n_{t} l_{t} }$ are radial matrix elements for $\ket{n_{a} l_{a}} \to \ket{n_{s} l_{s}} $ and $\ket{n_{b} l_{b}} \to \ket{n_{t} l_{t} }$ transitions, respectively. We calculated the radial matrix elements using a quasiclassical approximation~\cite{Kaulakys1995}.

The F\"{o}rster energy defect for channel $k$ is the difference between the energies of the initial collective state  $\ket{r_a r_b}$ and of the final state  $\ket{r_s r_t}$:
\be
\label{eq14}
\hbar\delta_k= [U(r_a)-U(r_s)]+[U(r_b)-U(r_t)].
\ee

We consider the F\"{o}rster resonance $\ket{60P_{3/2},m_1=3/2; \, 80P_{3/2}, m_2=3/2} \to \ket{59D_{5/2}, m'_1=5/2; \, 78D_{5/2}, m'_2=5/2}$ with  $\Delta M=\pm2$. Therefore we have chosen $\theta=\pi/2$. The dependence of the energies of the collective states $\ket{r_i r_j}$  with $M=1,2,3$ for the $\ket{60P;80P}\to \ket{59D; 78D}$ F\"{o}rster resonance on the electric field is shown in Fig.~\ref{Stark}(b). Here we selected $\ket{60P_{3/2};80P_{3/2}}$ as zero energy state in zero electric field. The energy of the collective state $\ket{60P_{3/2}, m_1=3/2; \, 80P_{3/2}, m_2=3/2}$ state is shown as thick dashed line in Fig.~\ref{Stark}(b) and the energy of the $\ket{59D_{5/2}, m'_1=5/2; \, 78D_{5/2}, m'_2=5/2}$ collective state is shown as thick solid line in Fig.~\ref{Stark}(b).

In the region $E<0.4$~V/cm we can approximate the energy shifts of the Rydberg states as 
\be
\label{eq15}
\delta_i \left(t\right)= -{\alpha_i^{(1)} } E-\alpha_i^{(2)}  E\left(t\right)^{2}.
\ee

The coefficients $\alpha_i^{(1)}$ and $\alpha_i^{(2)}$ , calculated for each Rydberg state by approximation of Stark diagrams, are listed in Table~I.

\begin{table}
\caption{Calculated polarizabilities of Rb Rydberg states for $E<0.4$~V/cm}
\begin{tabular*}{\columnwidth}{@{\extracolsep{\fill}}|c|c|c|c|} \hline 
State & $\left|m_{j} \right|$ &  $\alpha^{(1)} \left[\frac{{\rm MHz}}{\left({{\rm V} \mathord{\left/{\vphantom{{\rm V} {\rm cm}}}\right.\kern-\nulldelimiterspace} {\rm cm}} \right) } \right]$ & $\alpha^{(2)} \left[\frac{{\rm MHz}}{\left({{\rm V} \mathord{\left/{\vphantom{{\rm V} {\rm cm}}}\right.\kern-\nulldelimiterspace} {\rm cm}} \right)^{2} } \right]$ \\ \hline 
$\ket{60P_{3/2}} $ & 3/2 & 0.67 & 565 \\ \hline 
$\ket{80P_{3/2}} $ & 3/2 & 96.3  & 3957 \\ \hline 
$\ket{59D_{5/2}} $ & 5/2 &  -0.38  & 345  \\ \hline 
$\ket{59D_{5/2}} $ & 3/2 &  5.27  & 23.2  \\ \hline 
$\ket{78D_{5/2}} $ & 5/2 &  -62.8 & 2851 \\ \hline 
$\ket{78D_{3/2}} $ & 3/2 &  -91.8 & 2871 \\ \hline

\end{tabular*}
\end{table}

Radiofrequency-assisted F\"{o}rster resonances were extensively studied in our recent works~\cite{Tretyakov2014, Yakshina2016}. Using radiofrequency or microwave electric field, it is possible to induce the "inaccessible" resonances which cannot be observed by applying dc electric field only. The energy defect for the $\ket{60P_{3/2};80P_{3/2}} \to \ket{59D_{5/2}; 78D_{5/2}}$ channel in Rb atoms in zero electric field is $\delta_1^{(0)}/(2\pi)=-69.84$~MHz [see Fig.~\ref{Stark}(b)] and it only increases if the dc electric field is applied. However, this resonance can be induced using the rf electric field added. The calculated interaction energy for $\theta=\pi/2$ is $V_{1}/(2 \pi)=12140\,\rm MHz \,\mu m^3$.

The time-dependent electric field required for adiabatic passage of rf-assisted F\"{o}rster resonances is written as 

\be
\label{eq16}
E\left(t\right)=E_{dc} \left(t\right)+E_{V} \cos \left(\omega _{rf} t\right).
\ee

\noindent Here $E_{dc} \left(t\right)$ is a slowly-varying electric field which tunes the collective energy levels across the rf-assisted F\"{o}rster resonance, and $E_{V}$ is the amplitude of the rf field.

The time-averaged detuning for the electric field from Eq.~(\ref{eq16}) is written as 
\be
\label{eq17}
\left\langle \delta_k \left(t\right)\right\rangle =\delta_{k}^{(0)} -\alpha_k^{(1)} E_{dc} \left(t\right) -\alpha_k^{(2)}E_{dc} \left(t\right)^{2} -\frac{\alpha_k^{(2)} }{2} E_{V} ^{2}.
\ee
Here $\delta_k^{(0)}$ is the F\"{o}rster energy defect in zero electric field and the coefficients $\alpha_k^{(1)}$ and $\alpha_k^{(2)}$ are calculated using Table~I as the differences of the polarizabilities of  collective states $\ket{r_a r_b}$ and $\ket{r_s r_t}$ for the interaction channel $k$. The first-order rf-assisted F\"{o}rster resonance occurs, when $\left\langle \delta \right\rangle =\pm \omega _{\mathrm{rf}} $.

The initial collective state in our calculations are $\ket{60P_{3/2},m_1=3/2; \;80P_{3/2},m_2=3/2}$.  We have calculated the probability amplitudes of collective states of two-atom system driven by the Rydberg-Ryderg interactions in the time-dependent electric field. We have included in our numeric model 68 collective states for the $\ket{60P;80P}\to\ket{59D;78D}$ F\"{o}rster resonance with $\left|M\right|=1,3,5$, taking into account the fine structure and Stark sublevels of the Rydberg states. To find the time dependence of the dc electric field, we write the condition for the time-averaged detuning, following Eqs.~(\ref{eq3}), (\ref{eq10}) and (\ref{eq14}):
\be 
\label{eq18}
\left\langle \delta_m \left(t\right)\right\rangle =-\omega_{rf} +s_{1} \left(t-t_{m} \right)+s_{2} \left(t-t_{m} \right)^{5}.
\ee
We have chosen $s_1/(2\pi)=-3.1\; \rm MHz/\mu \rm s$, $s_2/(2\pi)=-80\; \rm MHz/\mu \rm s^5$, $t_1=-0.8 \;\mu \rm s$, $t_2=0.8 \;\mu \rm s$  and the total length of the double adiabatic passage $T=3.2\;\mu \rm s$.  The frequency of the rf field is $\omega_{\rm rf}/(2\pi)=220$~MHz and the amplitude $E_V=0.1$~V/cm. The critical field for the rf-induced F\"{o}rster resonance is 0.274~V/cm, and the position of this resonance is marked by the arrow in Fig.~\ref{Stark}(b).

\begin{center}
\begin{figure}[!t]
\center
\includegraphics[width=\columnwidth]{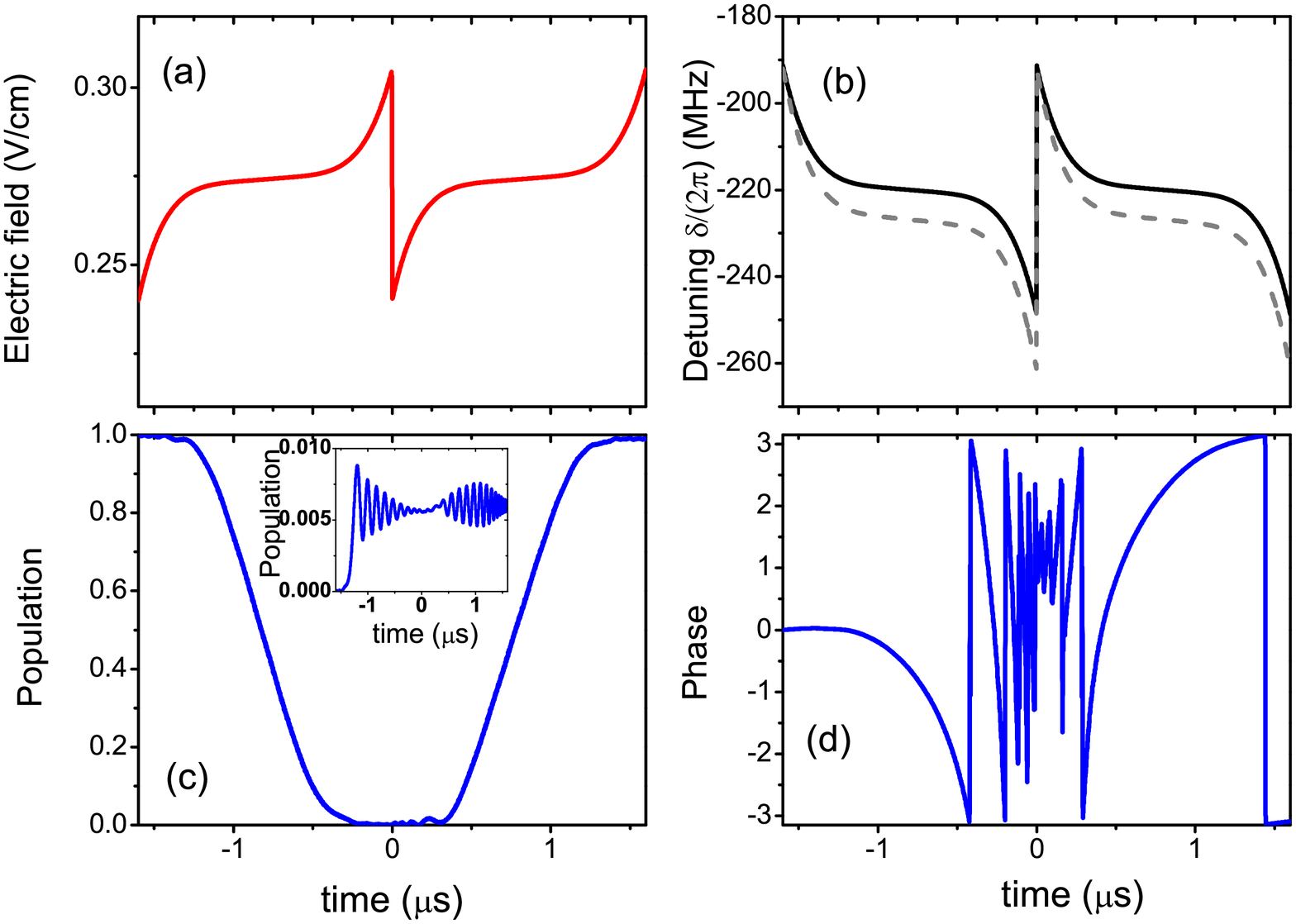}
\vspace{-.5cm}

\caption{
\label{Electric}
(Color online) (a) Time dependence of the dc electric field for Stark tuning of the rf-assisted $\ket{60P_{3/2},m_1=3/2; \, 80P_{3/2}, m_2=3/2} \to \ket{59D_{5/2}, m'_1=5/2; \, 78D_{5/2}, m'_2=5/2}$ F\"{o}rster resonance  with rf frequency of 220~MHz; (b) Time dependence of the F\"{o}rster the energy defects for the $\ket{60P_{3/2}, m_1=3/2; \, 80P_{3/2}, m_2=3/2} \to \ket{59D_{5/2}, m'_1=5/2; \, 78D_{5/2}, m'_2=5/2}$  interaction channel (solid line) and the $\ket{60P_{3/2},m_1=3/2; \, 80P_{3/2}, m_2=3/2} \to \ket{59D_{5/2}, m'_1=3/2; \, 78D_{3/2}, m'_2=3/2}$ interaction channel (dashed line);
(c) Time dependence of the population  of the collective state $\ket{60P_{3/2}, m_1=3/2; \, 80P_{3/2}, m_1=3/2;}$ and of the collective state $\ket{59D_{5/2}, m'_1=3/2; \, 78D_{3/2}, m'_2=3/2}$ (inset)  for $R=16.5\; \mu \rm m$. (d)  Time dependence of  phase  of the collective state $\ket{60P_{3/2}, m_1=3/2; \, 80P_{3/2}, m_1=3/2;}$  for $R=16.5\; \mu \rm m$.
}

\end{figure}
\end{center}

The time dependence of the dc electric field used for adiabatic passage of the rf-assisted F\"{o}rster resonances is shown in Fig.~\ref{Electric}(a). The time-averaged dependences of the energy defects for the $\ket{60P_{3/2}, m_1=3/2; \, 80P_{3/2}, m_2=3/2} \to \ket{59D_{5/2}, m'_1=5/2; \, 78D_{5/2}, m'_2=5/2}$ and the $\ket{60P_{3/2},m_1=3/2; \, 80P_{3/2}, m_2=3/2} \to \ket{59D_{5/2}, m'_1=3/2; \, 78D_{3/2}, m'_2=3/2}$ F\"{o}rster interaction channels are shown in Fig.~\ref{Electric}(b) as solid and dashed lines, respectively. These interaction channels are very close to each other, which is the main source of the infidelity of Bell states in our simulation. The F\"{o}rster energy defect for this resonance is $\delta_{2}^{(0)} /(2 \pi)=-46.11$~MHz and the interaction energy is $V_{2}/(2 \pi)=1613\,\rm MHz \,\mu m^3$.

\begin{center}
\begin{figure}[!t]
\center
\includegraphics[width=\columnwidth]{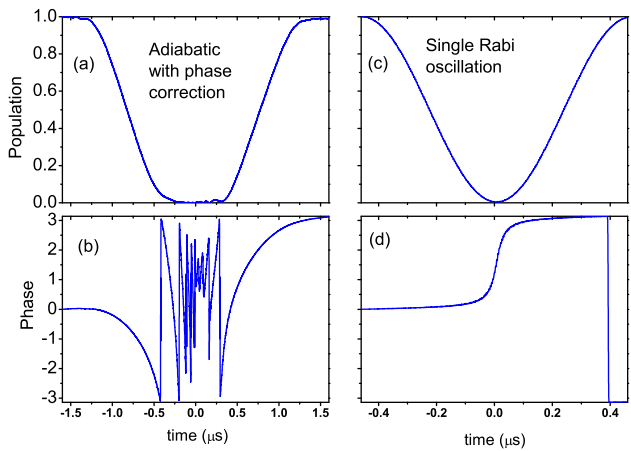}
\vspace{-.5cm}

\caption{
\label{Phase}
(Color online) Time dependence of the population  and phase  of the collective state $\ket{60P_{3/2}, m_1=3/2; \, 80P_{3/2}, m_1=3/2;}$  for $R=16.5\; \mu \rm m$. (a) Time dependence of the population during the double adiabatic passage with phase correction; (b) Time dependence of the phase during the double adiabatic passage with phase correction; (c) Time dependence of the population for coherent coupling; (d) Time dependence of the phase for coherent coupling.
}

\end{figure}
\end{center}

The numerically calculated time dependence of the population and phase of the collective $\ket{60P_{3/2}, m_1=3/2; \, 80P_{3/2}, m_1=3/2}$ state  is shown in Figs.~\ref{Electric}(c) and \ref{Electric}(d).  The distance between the atoms is fixed and taken to be $R=16.5\; \mu \rm m$. After the end of the double adiabatic passage the population returns back to the initial state with 97.9\% probability. Here the lifetimes of Rydberg states are not taken into account. The main source of error is the population leakage to the $\ket{59D_{5/2}, m'_1=3/2; \, 78D_{3/2}, m'_2=3/2}$ state during double adiabatic passage, as illustrated in the inset of Fig.~\ref{Electric}(c). Due to non-resonant interactions the phase of the initial state  $\ket{60P_{3/2}, m_1=3/2; \, 80P_{3/2}, m_1=3/2}$  after the end of the adiabatic passage is not exactly equal to $\pi$, as shown in Fig.~\ref{Electric}(d). Similarly to our previous work~\cite{Beterov2016a}, we correct this phase error by setting $t_{2} =0.7996\; \mu s$. The time dependence of the population and phase of the $\ket{60P_{3/2}, m_1=3/2; \, 80P_{3/2}, m_1=3/2}$  state with such correction is shown in Figs.~\ref{Phase}(a) and ~\ref{Phase}(b), respectively. 

For comparison, we also considered coherent coupling at the  rf-assisted F\"{o}rster resonance in the critical dc electric field of 274~mV/cm. Single Rabi-like oscillation occurs during the time interval $T=0.92\; \mu \rm s$, as shown in Fig.~\ref{Phase}(c).  The time dependence of the phase of the ground state is shown in Fig.~\ref{Phase}(d).The coherent coupling at F\"{o}rster resonance can be implemented at substantially shorter timescale than adiabatic passage, thus reducing the decoherence from finite lifetimes of Rydberg states. However, its drawback is that it requires precise control of the interatomic distance to avoid population and phase errors, which is hard to achieve.

\section{Generation of Bell states  using adiabatic passage}

\begin{center}
\begin{figure}[!t]
\center
\includegraphics[width=\columnwidth]{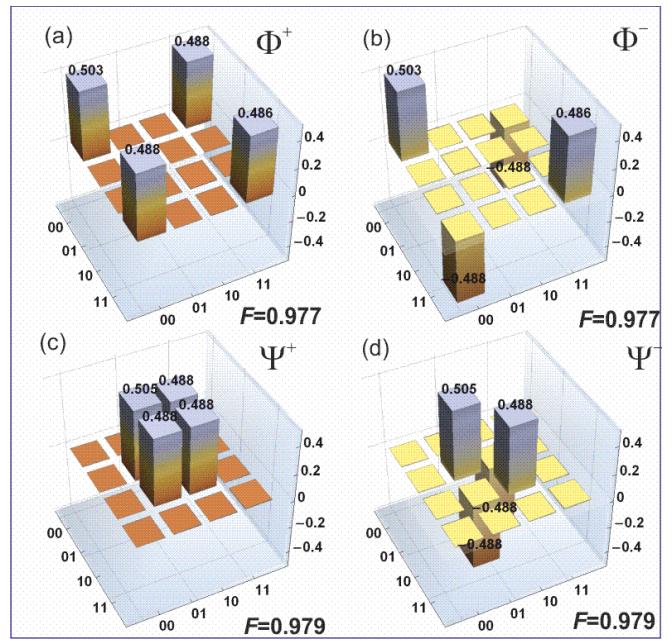}
\vspace{-.5cm}

\caption{
\label{Bell}
(Color online)  The numerically reconstructed density matrices of the (a) $\Phi ^{+} $, (b) $\Phi ^{-} $ , (c) $\Psi ^{+} $, (d) $\Psi ^{-} $ Bell states. 
}

\end{figure}
\end{center}

The entangled Bell states of a bipartite quantum system are defined as following~\cite{Bell1964, Nielsen2011}:

\bea
\label{eq21}
\Phi^+&=&\frac 1{\sqrt 2}\left(\ket{00} +\ket{11} \right), \nonumber\\ 
\Phi ^-&=&\frac 1{\sqrt 2}\left(\ket{00} -\ket{11} \right), \nonumber\\
\Psi ^+&=&\frac 1{\sqrt 2}\left(\ket{01} +\ket{10} \right), \nonumber\\
\Psi ^-&=&\frac 1{\sqrt 2}\left(\ket{01}-\ket{10} \right).
\eea

We have calculated the fidelities of the generated Bell states for Rb atoms taking into account finite lifetimes of Rydberg states~\cite{Beterov2009}. The scheme of generation of Bell states using a CZ gate is shown in Fig.~\ref{Scheme}(b). The Hadamard gate can be implemented using either microwave or Raman laser transitions between the  hyperfine sublevels $\ket{0}$, $\ket{1}$ of the ground state.  For simplicity, we consider an open system and neglect return of the population from Rydberg to the ground state due to spontaneous decay during adiabatic passage. The time dynamics of the probability amplitudes in the open system with finite lifetimes of the Rydberg state \textit{i} can be described by the time-dependent Schr\"{o}dinger equation by adding $-\gamma _i/2 $ to the right-hand side of each equation for the probability amplitude $c_{i}$~\cite{Shore2008}, where $\gamma_{i} $ is a decay rate of the Rydberg state.  

For simplicity, in our simulations we prepared the control qubit in the superposition state $\left(\ket{0} +\ket{1} \right)/\sqrt{2}$ or $\left(\ket{0} -\ket{1} \right)/\sqrt{2}$ which should appear after the Hadamard gate is applied to the control qubit as shown in Fig.~\ref{Scheme}(b). Then we applied the short microwave and laser pulses (with 10~ns duration)  to the control and target qubits to perform Hadamard gates on the target qubit and then to excite both atoms into the Rydberg state following the scheme of Fig.~\ref{Scheme}(b). We considered laser excitation in a zero electric field. We have taken into account the accumulation of the phase of the excited $60P_{3/2}$ and $80P_{3/2}$ states in the time-dependent electric field during adiabatic passage by adjusting the phase of the laser pulses which de-excite Rydberg states after the adiabatic passage is finished.

The numerically reconstructed density matrices for the Bell states are shown in Fig.~\ref{Bell}. We see that the  fidelity is  close to 97.7\% for all Bell states.

\begin{center}
\begin{figure}[!t]
\center
\includegraphics[width=\columnwidth]{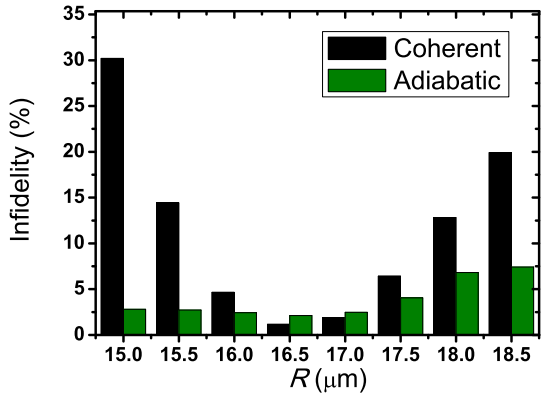}
\vspace{-.5cm}

\caption{
\label{Fidelity}
(Color online) Dependence of the infidelity of Bell states on interatomic distance.
}

\end{figure}
\end{center}

We have calculated also the fidelity of $\Psi^+$ Bell state for coherent coupling at F\"{o}rster resonance for the same interatomic distances to compare with the double adiabatic passage. The calculated dependence of the infidelity of generated Bell states is shown as a function of interatomic distance in Fig.~\ref{Fidelity}. We compare the infidelity of the Bell state generated using  by double adiabatic passage of the rf-assisted F\"{o}rster resonance with the fidelity of Bell state generated using coherent coupling at rf-assisted F\"{o}rster resonance. From Fig.~\ref{Fidelity} it is clear that double adiabatic passage reduces the sensitivity of the fidelity of  Bell states to the fluctuations of the interatomic distance compared with coherent coupling at rf-assisted F\"{o}rster resonance at a price of the reduced fidelity.

\section{Conclusion}

Implementation of high-fidelity two-qubit gates with Rydberg atoms is a challenging problem. In the schemes based on coherent coupling due to Rydberg-Rydberg interactions, off-resonant F\"{o}rster interactions and finite lifetimes of Rydberg states strongly limit the fidelity of the generated Bell states. RF-assisted F\"{o}rster resonances provide additional flexibility in search of well isolated F\"{o}rster resonances which are suitable for implementation of two-qubit gates and generation of Bell states with ultracold neutral atoms. We have identified such resonances for Rb atoms  and have shown that rf-assisted F\"{o}rster resonances allow coherent coupling and adiabatic rapid passage for population inversion. We have shown that the double adiabatic passage of rf-assisted F\"{o}rster resonances results in the deterministic phase shift and allows reducing the sensitivity of fidelity of Bell states to the fluctuations of the interatomic distance. Both coherent coupling and double adiabatic passage at rf-assisted  F\"{o}rster resonance can be used for generation of entanglement between the atoms at large interatomic distances where Rydberg blockade is cannot be obtained due to small interaction energy.  This can be especially useful for the quantum registers based on the dipole trap arrays obtained with microlenses arrays~\cite{Dumke2016}. Coherent coupling at F\"{o}rster resonance can be used, in principle, to achieve higher fidelity of two-qubit gates, but it requires precise control of the interatomic distance which can be experimentally challenging. Therefore we believe that the double adiabatic passage can be used to achieve reasonably high fidelity in experiments with Rydberg atoms.

\begin{acknowledgments}
This work was supported by the Russian Science Foundation Grant No. 16-12-00028 in the part of estimates of infidelity of Bell states, by RFBR Grants No. 17-02-00987 and 16-02-00383, by Novosibirsk State University and Russian Academy of Sciences. 
\end{acknowledgments}

%

\end{document}